\documentstyle[11pt,a4wide,epsfig]{article}
\begin{document}

\begin{titlepage}
\noindent
\begin{flushright}
December 1998
\end{flushright}

\vfill

\begin{center}
\noindent
{\huge\bf{Anomalous couplings of vector bosons and the decay $H
\longrightarrow \gamma \gamma$:  Dimensional regularization
versus momentum cutoff}}

\vspace{1cm}

{\large J. Novotn\'{y}{\footnote{e-mail: {\tt
Jiri.Novotny@mff.cuni.cz}}} and M. St\"{o}hr{\footnote{e-mail:
{\tt Miroslav.Stohr@mff.cuni.cz}}} \\}
{\it Nuclear Centre, Faculty of Mathematics and Physics, Charles University\\
V Hole\v{s}ovi\v{c}k\'{a}ch 2, Prague 8, Czech Republic\\}

\end{center}

\vfill

\begin{center}
{\bf Abstract}
\end{center}

As an illustration of general principles, the $W$-boson loop
contribution to the amplitude for the decay
$H\rightarrow \gamma \gamma $ is calculated within a
specific model for the effective lagrangian describing the anomalous
gauge boson couplings.
Different approaches to the dependence
of the result  on the scale of new physics $\Lambda$  are briefly discussed.


\vfill

\end{titlepage}

\newpage

\setcounter{footnote}{0}

\section{Introduction}

It is widely believed that the standard model (SM) of electroweak
interactions \cite{SM}, although so far in beautiful agreement with
experimental data, is not a final theory, but rather an effective low energy
theory\footnote{%
For a recent review and complete list of references on the application of
the effective lagrangian approach within the theory of the electroweak
interaction see e.g. \cite{Wudka} and references therein. For the general
principles of the effective lagrangians see also \cite{Georgi}.}, valid for
energies well below some scale $\Lambda $, which characterizes the onset of
a new physics. This scale could be e.g. the mass of yet unobserved heavy
particles present in the spectrum of the more fundamental theory.
Integrating out these high energy degrees of freedom, one ends up with an
effective lagrangian which describes only the interactions of the particles
belonging to the low energy part of the original spectrum. Such an effective
lagrangian should reflect the most general features of the original theory
such as various types of symmetries, modes of their realization, anomaly
matching conditions etc. On the other hand it is not constrained to be
renormalizable in the usual sense, i.e. it can (and in general it has to, in
order to be consistent) contain interaction terms with canonical dimension
larger then four. These are multiplied with coupling constants, proportional
to the inverse powers of some mass parameter related to the scale $\Lambda $%
, and therefore suppressed with respect to the renormalizable terms (the SM
should be understood within this framework as the lowest order part of the
complete effective lagrangian within the expansion in the powers of $\Lambda
^{-1}$ ).The low energy coupling constants (LEC) could be at least in
principle calculable (e.g. by integrating the heavy degrees of freedom order
by order in perturbation theory and imposing necessary matching conditions
at the threshold of new physics) provided the fundamental theory were known.
However in the most of applications of the effective lagrangian approach
this is not the case either because the explicit calculations are not
possible, or, as in the case of SM, the fundamental theory of the new
physics is not known. Rather one lets these parameters a priori undetermined
and treat them as a useful parameterization of the dynamics of the yet
unknown fundamental theory. Relaxing the constraint of the renormalizability
of the interactions does not mean at all the complete loss of predictivity,
which was the traditional argument for rejecting such theories. As a rule,
within the framework of the effective theories there are well defined
expansion prescriptions, which enable one to calculate loops and to absorb
the infinities in the finite number of renormalized LEC at each order. These
may be in principle measured experimentally and then used as an input for
other predictions. Going to higher orders in the expansion, the number of
LEC increases considerably, however their importance decreases because of
suppression by negative powers of $\Lambda $. One can adopt also another
point of view and use the measurement of various LEC as the experimental
tests of variants of the models of the new physics.

There are generally two different types of effective lagrangians
parametrizing the physics beyond the SM, corresponding to two different
scenarios of breaking the gauge symmetry $SU(2)_L\otimes U(1)_Y\rightarrow
U(1)_{em}$. The ''decoupling scenario'' assumes that the scale of the new
physics is much larger than the electroweak symmetry breaking scale, $%
\Lambda \gg v$ , (here $v$ is the vacuum expectation value of the Higgs
doublet), the $SU(2)_L\otimes U(1)_Y$ symmetry is then linearly realized and
the low energy spectrum is identical with that of the SM, including the
Higgs particle, which is supposed to be relatively light. The new
nonrenormalizable interactions are organized according to the increasing
canonical dimension:
\begin{equation}
{\cal L}_{eff}={\cal L}_{SM}+\sum_i\frac{f_i^{(6)}}{\Lambda ^2}{\cal O}%
_i^{(6)}+\sum_i\frac{f_i^{(8)}}{\Lambda ^4}{\cal O}_i^{(8)}+\ldots
\label{linear}
\end{equation}
and includes all operators of dimension 6, 8, ... invariant with respect to
the $SU(2)_L\otimes U(1)_Y$.

The ''nondecoupling scenario'' on the other hand corresponds to the case $%
\Lambda \approx 4\pi v$. The new physics is related to the symmetry breaking
sector of the SM, the gauge symmetry is realized nonlinearly and the
effective lagrangian reflects the dynamics of the would-be Goldstone bosons
eaten by the gauge bosons (and via the equivalence theorem \cite{ET}, it is
related to the dynamics of the longitudinal components of $W^{\pm }$ and $Z$%
). The most economic form of such an effective lagrangian for the symmetry
breaking sector is given in the form of the (gauged) nonlinear $\sigma $
model organized as a derivative expansion
\begin{equation}
{\cal L}_{eff}^{SB}=\frac{v^2}4{\rm tr}(D_\mu U^{+}D^\mu U)+\ldots ,
\label{chiral}
\end{equation}
(where $U=\exp ({\rm i}\xi ^a\tau ^a/v)\in SU(2)$, $\tau ^a$ are Pauli
matrices and $\xi ^a$ are the would-be Goldstone boson fields, $D_\mu
U=\partial _\mu U-g\widehat{W}_\mu U-({\rm i}g^{\prime }/2)B_\mu U$ and $%
\widehat{W}_\mu =(1/2{\rm i})W_\mu ^a\tau ^a$); the ellipses here mean terms
with four and more derivatives and/or gauge fields, invariant with respect
to the local $SU(2)_L\otimes U(1)_Y$ gauge transformation. This has for the $%
U$ field the following form
\[
U\rightarrow \exp ({\rm i}\alpha _L^a\frac{\tau ^a}2)U\exp (-{\rm i}\alpha _Y%
\frac{\tau ^3}2).
\]
In order to preserve the $\rho $ parameter to be close to one, the
additional (global) symmetries are often imposed. The example of such a
symmetry is the custodial $SU(2)_c$ symmetry, which is introduced as the
unbroken subgroup of the symmetry breaking pattern $SU(2)_L\otimes
SU(2)_R\rightarrow SU(2)_c$, completely analogous to the pattern of chiral
symmetry breaking of QCD with two light quarks \cite{chpt}. The field $U$
transforms under $SU(2)_L\otimes SU(2)_R$ according to

\[
U\rightarrow \exp ({\rm i}\alpha _L^a\frac{\tau ^a}2)U\exp (-{\rm i}\alpha
_R^b\frac{\tau ^b}2).
\]
and the custodial symmetry corresponds to the diagonal subgroup $\alpha
_L^a=-\alpha _R^a$. Of course, gauging then the $SU(2)_L\otimes U(1)_Y$
subgroup means the explicit breaking of the $SU(2)_c$ by the terms which
vanish for $g^{^{\prime }}\rightarrow 0$. Note also, that there is no Higgs
field included in this type of effective lagrangian. However, also in this
case it is possible to extend the model to account for such a type of
particle \cite{Feruglio} adding to the lagrangian additional terms
containing $SU(2)_L\otimes U(1)_Y$ invariant field $H$:
\begin{eqnarray}
{\cal L}_{eff}^H &=&\frac 12\partial _\mu H\partial ^\mu H-V(H)+a\left(
\frac Hv\right) \frac{v^2}4{\rm tr}(D_\mu U^{+}D^\mu U)  \nonumber \\
&&\ \ \ +b\left( \frac Hv\right) ^2\frac{v^2}4{\rm tr}(D_\mu U^{+}D^\mu
U)+\ldots ,  \label{nonlinear Higgs}
\end{eqnarray}
the ellipses stand for terms of higher order as well as for the interaction
terms of $H$ field with SM fermions. The SM Higgs is recovered for
\begin{eqnarray}
V(H) &=&\frac{m_H^2}{8v^2}((H+v)^2-v^2)^2  \nonumber \\
a &=&2,\ b=1.  \label{SM Higgs}
\end{eqnarray}

In the unitary gauge, both these scenarios lead to the $U(1)_{em}$ invariant
effective lagrangian with anomalous couplings. E.g. in the gauge boson
sector, there are besides other contributions the following triple gauge
boson couplings, which are usually written in the form of phenomenological
lagrangian (we have omitted possible C or CP violating terms)\footnote{%
In fact, in such a phenomenological lagrangian the coupling constant should
be understood in general as $q$ dependent formfactors.}

\begin{eqnarray}
{\cal L}_{eff}^{WWV} &=&\sum_{V=\gamma ,Z}g_V[{\rm i}g_1^V(W_{\mu \nu
}^{+}W^\mu V^\nu -W_{\mu \nu }W^{+\mu }V^\nu )+{\rm i}\kappa _VW_\mu
^{+}W_\nu V^{\mu \nu } \\
&&\ \ \ +{\rm i}\frac{\lambda _V}{m_W^2}W_{\nu \mu }^{+}W_\lambda ^\mu
V^{\lambda \nu }],  \label{TGV}
\end{eqnarray}
where $W_{\mu \nu }=\partial _\mu W_\nu -\partial _\nu W_\mu $and
analogously for $V$, $g_\gamma =-e$ and $g_Z=-g\cos \theta _W$. Note, that
within the SM, $g_1^V=\kappa _V=1$, $\lambda _V=0$. The constants $\kappa
_\gamma $ and $\lambda _\gamma $ can be interpreted in terms of the
anomalous magnetic and quadrupole moment of the $W$ bosons, $g_1^V$
corresponds to the gauge $U(1)$ charges of the $W$ in the the units of $%
g_\gamma $ an $g_Z$. Recent constraints on these couplings \cite{experiment}
come from the studies of $W\gamma $ events at Fermilab Tevatron; the CDF and
D0 results are $-1.6<\Delta \kappa _\gamma <1.8$ and $-0.6<\lambda _\gamma
<0.6$. Direct measurement will be also available at LEP2 \cite{LEP2}.

\section{Divergences within the effective theory and the dependence on the
scale of the new physics}

The $U(1)_{em}$ invariant phenomenological lagrangians reviewed in the
previous section were often used in various calculations and treated as
specific models of the deviations from the SM physics. There was certain
controversy in the literature concerning the treatment of the divergences,
which appear in the loops with anomalous vertices. Some of the authors used
in one or another way the momentum cutoff at the scale of new physics $%
\Lambda _{cutoff}=\Lambda _{NP}=\Lambda $. Within this approach, which seems
to be very physically illuminating, the principle \cite{Burgess} was used,
that the divergent graph cut off at the scale where the effective theory
looses its validity due the onset of the new physics gives a lower bound to
the actual value of the graph in the full theory. I.e., in practise, only
the most divergent contribution of the loop integral to the given amplitude
was kept and used as the estimate of the dependence of the full theory
amplitude on the scale of the new physics.The appearance of the divergences
was therefore interpreted as an indication, that the process under
consideration is strongly sensitive to the scale of the new physics.

This approach has been criticized (see e.g. the paper \cite{Burgess})
because it is in conflict with the decoupling theorems and gives ambiguous
results. Moreover, as usual, it highly overestimates the dependence of the
result on the scale\footnote{$\Lambda $ appears in the results calculated
within this approach in a positive power (or logarithm) coming from the
power counting order of the divergent loops.} $\Lambda $ .

However, there are explicit examples discussed e.g. in \cite{Burgess}
illustrating that the momentum cutoff could yield correct results\footnote{%
The result depends strongly on the choice of the field variables. The
correct dependence on the scale of the new physics can be read off from the
cutoff dependence of the loops only using the ''good'' variables. However,
there is no clear criteria how to decide, whether the choice of variables is
''good'' or ''bad''.}. As usual, it is not possible to resolve whether this
is the case before performing explicit calculations.

In this paper we would like to briefly illustrate this general situation by
using an explicit example of the $W$-boson loops contribution to the process
$H\rightarrow \gamma \gamma $ calculated within simple $U(1)_{em}$ invariant
phenomenological lagrangian in the unitary gauge using two regularization
schemes. We will show, that there are differences between the momentum
cutoff approach and another approach based on dimensional regularization
with minimal subtractions, which was advocated in \cite{Burgess} and which
should be accepted as the procedure giving the correct answer. We will
therefore conclude, that the process $H\rightarrow \gamma \gamma $ is not
reckoned among the cases, which could be treated correctly within the
momentum cutoff prescription in the unitary gauge.

The paper is organized as follows. In Section 2 we introduce a specific
model of the effective lagrangian with anomalous gauge boson couplings and
present the results for the amplitude $H\rightarrow \gamma \gamma $
calculated within the two above mentioned schemes. In Section 3 we discuss
the result of the calculations from the general point of view which was
sketched in the Introduction. The comparison of the two approaches with
respect to the dependence of the decay amplitude on the scale of the new
physics and the conclusions are presented in Section 4. The details of the
calculation of the amplitude are postponed to the Appendix.

\section{$H\rightarrow \gamma \gamma $ within the effective lagrangian
approach - a specific model}

The theoretical concentration on the decay $H\rightarrow \gamma \gamma $ in
the recent literature \cite{Ellis} -- \cite{Hagiwara} is motivated by the
fact that this rare decay mode could serve as the main source of the
experimental signal for the Higgs particle with the mass within the lower
part of the intermediate mass range $m_Z<m_H<2m_W$ on hadron colliders (e.g.
LHC).

Within the Standard Model, this decay channel of the Higgs boson is
described at lowest order by a sum of one-loop Feynman diagrams, this sum is
ultraviolet finite (the reason is that there is no tree-level $H\gamma
\gamma $ interaction in the SM lagrangian; SM is a renormalizable theory, so
that the $H\gamma \gamma $ counterterms cannot be present). As a result, one
gets for the decay amplitude an expression of the following form \cite{Ellis}%
, the tensor structure of which reflects the Lorentz invariance and $U(1)_{em%
{\ }}$gauge invariance:

\begin{equation}
{\cal M}(H\rightarrow \gamma \gamma )=\frac{e^2g}{16\pi ^2}\frac
1{m_W}\;\varepsilon _\mu ^{\star }\left( k\right) \varepsilon _\nu ^{\star
}\left( l\right) \left[ \frac 12m_H^2g^{\mu \nu }-k^\nu l^\mu \right]
\;\sum_{i=s,f,g}N_{c_i}e_i^2F_i.  \label{SM amplitude}
\end{equation}
The decay rate is then

\begin{eqnarray}
\Gamma (H\rightarrow \gamma \gamma )=\frac{\alpha ^3}{128\pi ^2\sin ^2\theta
_W}\frac{m_H^3}{m_W^2}\left| \sum_{i=s,f,g}N_{c_i}e_i^2F_i\right| ^2.
\label{SM result}
\end{eqnarray}
Here

\begin{eqnarray}
F_s &=&\tau _s\left( 1-\tau _sI^2\right)  \nonumber  \label{eq12} \\
F_f &=&-2\tau _f\left[ 1+\left( 1-\tau _f\right) I^2\right]  \nonumber \\
F_g &=&2+3\tau _g+3\tau _g\left( 2-\tau _g\right) I^2  \label{F}
\end{eqnarray}
are the contributions of the scalar, fermion and vector boson loops resp.,
\begin{eqnarray}
\tau _i &=&4\left( \frac{m_i}{m_H}\right) ^2  \nonumber  \label{eq13} \\
N_{c_i} &=&1\;\;\mbox{for}\;i=\mbox{leptons, scalars and vector
bosons} \\
N_{c_i} &=&3\;\;\mbox{for}\;i=\mbox{quarks},  \nonumber
\end{eqnarray}
$e_i$ is electric charge of the loop particle in units of $e$, and
\begin{equation}
I=\left\{ \matrix{\arctan \frac{1}{\sqrt{\tau-1}} & ,\tau>1 \cr \frac{1}{2}
\left[ \pi + i \ln \left| \frac{1+\sqrt{1-\tau}}{1-\sqrt{1-\tau}} \right|
\right] & ,\tau<1}\right. .  \label{eq14}
\end{equation}

It was believed that this process could be significantly influenced by
possible deviations from the Standard Model, mainly in the gauge boson
sector. To parametrize the physics beyond the Standard Model one can employ
the effective lagrangian approach sketched above. This was done in several
papers (see e.g. \cite{Konig} -- \cite{Hagiwara}); the specific forms of the
used effective lagrangians vary among different authors. Since the
additional effective couplings are generally of a non-renormalizable type,
the resulting decay amplitude is, in contrast to the SM, UV divergent, so
that the loop integrals have to be regularized to obtain reasonable
predictions. This is another point, at which the above mentioned papers
differ.

In order to illustrate this general situation we would like to present here
a complementary alternative to the treatment contained in the work \cite
{Konig}, in which the $W$-boson contribution to the decay width $%
H\rightarrow \gamma \gamma $ was calculated within the effective lagrangian
approach. Using the same form of an effective lagrangian as in \cite{Konig},
we would like to demonstrate the differences resulting from a different
cutoff prescription. We will also shortly comment on the interpretation of
the output of the explicit calculations.

Let us first briefly review the main results of \cite{Konig} and the way
they were obtained. As the phenomenological lagrangian describing the gauge
boson sector it was used the $U(1)_{em}$ invariant lagrangian ${\cal {L}}$
introduced originally in \cite{Aronson} (cf. also \cite{Grifols et al} ),
conserving $C$ and $P$ separately. In the same notation as in\cite{Konig},
this phenomenological lagrangian is given in the unitary gauge as a sum of
three terms
\begin{eqnarray}
{\cal {L}}={\cal {L}}_1+{\cal {L}}_2+{\cal {L}}_3,  \label{lagrangian}
\end{eqnarray}

where

\begin{eqnarray}
{\cal {L}}_1 &=&-\frac 12\widehat{G}_{\mu \nu }^{\dagger }\widehat{G}^{\mu
\nu }+m_W^2W_\mu ^{\dagger }W^\mu +\frac 12m_Z^2Z_\mu Z^\mu  \label{l1} \\
{\cal {L}}_2 &=&-\frac 14\sum_{V=\gamma Z}\widehat{F}_{\mu \nu }^{(V)}%
\widehat{F}^{(V)\mu \nu }  \label{l2} \\
{\cal {L}}_3 &=&\sum_{V=\gamma Z}\frac{ig_V\lambda _V}{m_W^2}\widehat{F}%
_{\mu \nu }^{(V)}\widehat{G}^{\dagger \mu \rho }\widehat{G}_\rho ^{\;\;\nu }.
\label{l3}
\end{eqnarray}
Here
\begin{eqnarray}
\widehat{G}_{\mu \nu } &=&\left( \partial _\mu -ig_\gamma A_\mu -ig_ZZ_\mu
\right) W_\nu -\left( \partial _\nu -ig_\gamma A_\nu -ig_ZZ_\nu \right) W_\mu
\nonumber  \label{eq5} \\
\widehat{F}_{\mu \nu }^{(V)} &=&F_{\mu \nu }^{(V)}+ig_V\kappa _V\left( W_\mu
^{\dagger }W_\nu -W_\nu ^{\dagger }W_\mu \right)  \nonumber \\
F_{\mu \nu }^{(V)} &=&\partial _\mu V_\nu -\partial _\nu V_\mu  \nonumber \\
g_\gamma &=&e  \nonumber \\
g_Z &=&g\cos \theta _W.  \label{structures}
\end{eqnarray}

Let us note, that the triple gauge boson couplings correspond to (\ref{TGV})
with $g_1^\gamma =g_1^Z=1$. The relation $g_1^\gamma =1$ is quite natural,
because it expresses the conservation of the electric charge of $W$ bosons,
i.e. the $WW\gamma $ couplings derived from ${\cal {L}}$ are independent
linear combination of the most general $U(1)_{em}$ terms conserving $C$ and $%
P$. This is, however, not true for the $WW\gamma \gamma $ couplings; here
the possible interaction terms are not parametrized by independent LEC but
multiplied by very specific combinations of $\kappa _\gamma $ and $\lambda
_\gamma .$ For $\kappa _V=1$ and $\lambda _V=0$, the lagrangian ${\cal {L}}$
reduces to the lagrangian of SM in the unitary gauge.

In addition to the above lagrangian of the gauge sector, the Higgs boson was
included with the standard couplings to $W$ boson pair,

\begin{equation}
{\cal {L}}_{WWH}=gm_WW_\mu ^{+}W^\mu H.  \label{HWW}
\end{equation}

We can think about the above $U(1)_{em}$ invariant phenomenological
lagrangian as being produced by fixing the unitary gauge in the $%
SU(2)_L\otimes U(1)_Y$ invariant lagrangian within both decoupling and
nondecoupling scenarios. Within the nondecoupling scenario, there are three
operators of the order ${\cal O}(p^4)$ in the derivative expansion and three
operators of the order ${\cal O}(p^6)$ needed to reproduce the structure of
the lagrangian (\ref{lagrangian}) in the unitary gauge, namely\footnote{%
In fact, the operator {\rm tr}$(T[V_\mu ,V_\nu ])${\rm tr}$(T[V^\mu ,V^\nu
]) $ can be expressed in terms of the following elements of the operator
basis introduced in \cite{Longhitano}:
\begin{eqnarray*}
{\rm tr}(T[V_\mu ,V_\nu ]){\rm tr}(T[V^\mu ,V^\nu ]) &=&4([{\rm tr}(V_\mu
V_\nu )]^2-[{\rm tr}(V_\mu V^\mu )]^2-{\rm tr}(V_\mu V_\nu ){\rm tr}(TV^\mu )%
{\rm tr}(TV^\nu ) \\
&+&{\rm tr}(V_\mu V^\mu ){\rm tr}(TV_\nu ){\rm tr}(TV^\nu )).
\end{eqnarray*}
\par
The same can be done with the operator {\rm tr}$(T[V_\nu ,V_\mu ]){\rm tr}(T%
\widehat{W}^{\mu \rho }\widehat{W}_\rho ^{\;\;\nu })$, here
\par
\begin{eqnarray*}
{\rm tr}(T[V_\nu ,V_\mu ]){\rm tr}(T\widehat{W}^{\mu \rho }\widehat{W}_\rho
^{\;\;\nu }) &=&{\rm tr}(V_\nu V_\mu ){\rm tr}(\widehat{W}^{\mu \rho }%
\widehat{W}_\rho ^{\;\;\nu })-{\rm tr}(V_\mu \widehat{W}^{\mu \rho }){\rm tr}%
(\widehat{W}_\rho ^{\;\;\nu }V_\nu ) \\
&&+2{\rm tr}(TV_\mu ){\rm tr}(T\widehat{W}^{\mu \rho }){\rm tr}(\widehat{W}%
_\rho ^{\;\;\nu }V_\nu )-2{\rm tr}(T\widehat{W}_\rho ^{\;\;\nu }){\rm tr}%
(TV_\mu ){\rm tr}(V_\nu \widehat{W}^{\mu \rho })
\end{eqnarray*}
\par
\label{footnote}}

\begin{eqnarray}
{\cal L} &=&\frac 12{\rm tr}\widehat{W}_{\mu \nu }\widehat{W}^{\mu \nu
}-\frac 14B_{\mu \nu }B^{\mu \nu }+\frac{v^2}4{\rm tr}(D_\mu U^{+}D^\mu U)
\nonumber \\
&&+\frac 1{4g^2}(s_W^2\Delta \kappa _\gamma +c_W^2\Delta \kappa _Z){\rm tr}%
(T[V_\mu ,V_\nu ]){\rm tr}(T[V^\mu ,V^\nu ])  \nonumber \\
&&-\frac 1{2g}(s_W^2\Delta \kappa _\gamma +c_W^2\Delta \kappa _Z){\rm tr}(T%
\widehat{W}_{\mu \nu }){\rm tr}(T[V^\mu ,V^\nu ])  \nonumber \\
&&+\frac{{\rm i}}{2g}s_Wc_W(\Delta \kappa _\gamma -\Delta \kappa _Z)B_{\mu
\nu }{\rm tr}(T[V^\mu ,V^\nu ])  \nonumber \\
&&+\frac 23\frac g{m_W^2}(s_W^2\lambda _\gamma +c_W^2\lambda _Z){\rm tr}(%
\widehat{W}_{\nu \mu }\widehat{W}^{\mu \rho }\widehat{W}_\rho ^{\ \;\nu })
\nonumber \\
&&-{\rm i}\frac g{m_W^2}s_Wc_W(\lambda _\gamma -\lambda _Z)B_{\nu \mu }{\rm %
tr}(T\widehat{W}^{\mu \rho }\widehat{W}_\rho ^{\;\;\nu })  \nonumber \\
&&-\frac 1{m_W^2}(s_W\lambda _\gamma \Delta \kappa _\gamma +c_W\lambda
_Z\Delta \kappa _Z){\rm tr}(T[V_\nu ,V_\mu ]){\rm tr}(T\widehat{W}^{\mu \rho
}\widehat{W}_\rho ^{\;\;\nu }),  \label{nondecoupling}
\end{eqnarray}

where

\begin{eqnarray*}
T &=&U\tau ^3U^{+},\;V_\mu =(D_\mu U)U^{+},\;D_\mu U=\partial _\mu U-g%
\widehat{W}_\mu U+g^{^{\prime }}U\widehat{B}_\mu , \\
\widehat{W}_\mu &=&\frac 1{2{\rm i}}W_\mu ^a\tau ^a,\;\widehat{B}_\mu =\frac
1{2{\rm i}}B_\mu \tau ^3,\; \\
\widehat{W}_{\mu \nu } &=&\partial _\mu \widehat{W}_\nu -\partial _\nu
\widehat{W}_\mu -g[\widehat{W}_\mu ,\widehat{W}_\nu ],B_{\mu \nu }=\partial
_\mu B_\nu -\partial _\nu B_\mu
\end{eqnarray*}
and $\Delta \kappa _V=\kappa _V-1$. The lagrangian (\ref{HWW}) can be
obtained in the same way from ${\cal O}(p^2)$ term

\[
{\cal {L}}_{WWH}=\frac{v^2}2\left( \frac Hv\right) {\rm tr}(D_\mu U^{+}D^\mu
U).
\]

Within the decoupling scenario, the same anomalous gauge boson couplings are
reproduced by the lagrangian with two dimension 6, three dimension 8 and one
dimension 10 operators (cf. e.g. \cite{Renard}) :

\begin{eqnarray}
{\cal L} &=&\frac 12{\rm tr}\widehat{W}_{\mu \nu }\widehat{W}^{\mu \nu
}-\frac 14B_{\mu \nu }B^{\mu \nu }+D_\mu \Phi ^{+}D^\mu \Phi  \nonumber \\
&&+\frac{4{\rm i}}{gv^2}s_Wc_W(\Delta \kappa _\gamma -\Delta \kappa
_Z)B_{\mu \nu }D^\mu \Phi ^{+}D^\nu \Phi  \nonumber \\
&&+\frac 23\frac g{m_W^2}(s_W^2\lambda _\gamma +c_W^2\lambda _Z){\rm tr}(%
\widehat{W}_{\nu \mu }\widehat{W}^{\mu \rho }\widehat{W}_\rho ^{\ \;\nu })
\nonumber \\
&&+\frac{16}{gv^4}(s_W^2\Delta \kappa _\gamma +c_W^2\Delta \kappa _Z)\Phi
^{+}\widehat{W}_{\nu \mu }\Phi D^\mu \Phi ^{+}D^\nu \Phi  \nonumber \\
&&+\frac 4{gv^4}(s_W^2\Delta \kappa _\gamma +c_W^2\Delta \kappa _Z)[D^\mu
\Phi ^{+}D^\nu \Phi -D^\nu \Phi ^{+}D^\nu \Phi ]^2  \nonumber \\
&&+\frac{4{\rm i}}{m_W^2v^2}s_Wc_W(\lambda _\gamma -\lambda _Z)B_{\nu \mu
}\Phi ^{+}\widehat{W}^{\mu \rho }\widehat{W}_\rho ^{\ \;\nu }\Phi  \nonumber
\\
&&+\frac{16}{m_W^2v^4}(s_W\lambda _\gamma \Delta \kappa _\gamma +c_W\lambda
_Z\Delta \kappa _Z)  \nonumber \\
&&\times (D^\mu \Phi ^{+}D^\nu \Phi -D^\nu \Phi ^{+}D^\nu \Phi )\Phi ^{+}%
\widehat{W}^{\mu \rho }\widehat{W}_\rho ^{\ \;\nu }\Phi ,  \label{decoupling}
\end{eqnarray}
Here $\Phi $ is the SM Higgs doublet , $s_W=\sin \theta _W,$ $c_W=\cos
\theta _W$ and $D^\nu \Phi =\partial _\mu \Phi -g\widehat{W}_\mu \Phi +{\rm i%
}g^{^{\prime }}/2B_\mu \Phi .$ The standard Higgs boson coupling (\ref{HWW})
stems now from the third term of (\ref{decoupling}). Such a lagrangian
produces, however, also anomalous Higgs boson couplings, which were not
considered in \cite{Konig}.

Both the nondecoupling and decoupling interpretations of the origin of the
phenomenological lagrangian (\ref{lagrangian}) illustrate again the fact,
that this lagrangian is incomplete in the sense, that it does not contain
all the independent linear combinations of the full set of the operators up
to a given dimension or number of (covariant) derivatives\footnote{%
Let us also note that the suppression of the higher dimension or higher
order operators does not correspond to the usual factors of the type $%
1/\Lambda ^2$ for the decoupling scenario and $1/(4\pi v)$ for the
nondecoupling scenario, but rather to the factor $g^2/m_W^2\sim 1/v^2$ . The
reason is, that the contributions of all the operators to the parameters $%
\kappa _\gamma $ and $\lambda _\gamma $ should have the same order of
magnitude in order to reproduce the lagrangian \ref{lagrangian}.}.
Nevertheless, we use it here as it stands for the illustrative purposes,
mainly because of the relative calculational simplicity.

In the paper \cite{Konig}, only the $WWH$, $WW\gamma $ and $WW\gamma \gamma $
vertices derived from the above lagrangian (\ref{lagrangian}) were used in
the calculation of the $W$ -boson contribution to the $H\rightarrow \gamma
\gamma $ decay width. The calculation was performed in the unitary gauge.
There are two types of Feynman diagrams, namely the triangle (and
corresponding cross diagram) and the tadpole; both of them, when regulated
using momentum cutoff, lead to the quadratic divergent loop integrals. The
sum of these diagrams remain UV divergent unless $\kappa _\gamma =1$ and $%
\lambda _\gamma =0$, as it was explained above. The result given in \cite
{Konig} was presented in the form of the sum of the (cutoff independent) SM
expression and the (cutoff dependent) correction. The latter was identified
with the quadratic divergence of the diagrams with at least one anomalous
vertex proportional $\kappa _\gamma -1$ and/or $\lambda _\gamma .$ I.e.

\begin{eqnarray}
F_g &=&2+3\tau _g+3\tau _g\left( 2-\tau _g\right) I^2  \nonumber \\
&&\ +\left( \frac \Lambda {m_W}\right) ^2\left( 3\Delta \kappa _\gamma
-4\lambda _\gamma +\lambda _\gamma ^2+\frac 12\Delta \kappa _\gamma
^2\right) ,  \label{FKonig}
\end{eqnarray}
and the cutoff of the loop momentum $\Lambda $ was interpreted as the scale
where the new physics comes in. Because the divergences are local, within
this approach the effect of the loops with anomalous gauge boson coupling is
equivalent to some direct $H\gamma \gamma $ interaction vertex. It is not
difficult to see, that it corresponds to the vertex of the type
\[
{\cal L}_{H\gamma \gamma }=G_{eff}e^2\left( \frac Hv\right) F_{\mu \nu
}F^{\mu \nu },
\]
where the effective coupling constant reads
\begin{equation}
G_{eff}=\frac 1{(4\pi )^2}\left( \frac \Lambda {m_W}\right) ^2\left( 3\Delta
\kappa _\gamma -4\lambda _\gamma +\lambda _\gamma ^2+\frac 12\Delta \kappa
_\gamma ^2\right) .  \label{treecutoff}
\end{equation}

As an alternative to this, let us present here the result of our calculation
(the details of the calculation can be found in the Appendix) of the same
quantity $F_g$ within dimensional regularization and $\overline{MS}$
subtraction scheme. Such a treatment was advocated in \cite{Burgess}. The
result can be split up to the finite and divergent parts in the following
way
\begin{equation}
F_g=F_g^{fin}+F_g^{div},  \label{eq14_2}
\end{equation}
where the finite part is
\begin{eqnarray}
F_g^{fin} &=&3-{{\kappa _\gamma }^2+3\,\tau _g}+2\,\lambda _\gamma
+6\,\kappa _\gamma \,\lambda _\gamma -5\,{{\lambda _\gamma }^2}  \nonumber
\label{eq15} \\
&&\ {+\biggl( -1+2\,\kappa _\gamma -{{\kappa _\gamma }^2}+2\,\lambda _\gamma
-2\,\kappa _\gamma \,\lambda _\gamma -{{\lambda _\gamma }^2}}  \nonumber \\
&&\ {+\left( 3+2\,\kappa _\gamma +{{\kappa _\gamma }^2}-2\,\lambda _\gamma
-2\,\kappa _\gamma \,\lambda _\gamma +{{\lambda _\gamma }^2}\right) \,\tau
_g-3\,{{\tau _g}^2}\biggr) \,{I^2}}  \nonumber \\
&&\ +({{-3-2\,\kappa _\gamma +5\,{{\kappa _\gamma }^2}-2\,\lambda _\gamma
+2\,\kappa _\gamma \,\lambda _\gamma +\frac{25}3\,{{\lambda _\gamma }^2}}%
)\frac 1\tau _g}  \nonumber \\
&&\ {{+{{\Bigl(8-4\,\kappa _\gamma -4\,{{\kappa _\gamma }^2}+8\,\lambda
_\gamma +8\,\kappa _\gamma \,\lambda _\gamma -14\,{{\lambda _\gamma }^2}}}}}
\nonumber \\
&&\ +{{(-4+4\,{{\kappa _\gamma }^2}+8\,{{\lambda _\gamma }^2})}\frac 1\tau _g%
}  \nonumber \\
&&\ +{\ \left( -4+4\,\kappa _\gamma -8\,\lambda _\gamma -8\,\kappa _\gamma
\,\lambda _\gamma +6\,{{\lambda _\gamma }^2}\right) \,\tau _g\Bigr)\,\frac IJ%
}  \nonumber \\
&&\ +\left( -8\,\lambda _\gamma -4\,\kappa _\gamma \,\lambda _\gamma +6\,{{%
\lambda _\gamma }^2}+{{(2-2\,{{\kappa _\gamma }^2}-4\,{{\lambda _\gamma }^2})%
}\frac 1\tau _g}\right) \,\ln \left( \frac{m_W{^2}}{{{{\mu }^2}}}\right)
\nonumber \\
&&  \label{Ffinite}
\end{eqnarray}
and the divergent part is
\begin{eqnarray}
F_g^{div}{} &{}&{=}\left( -8\,\lambda _\gamma -4\,\kappa _\gamma \,\lambda
_\gamma +6\,{{\lambda _\gamma }^2}+{{(2-2\,{{\kappa _\gamma }^2}-4\,{{%
\lambda _\gamma }^2})}\frac 1\tau _g}\right)  \nonumber \\
&&\ \times {\,\left( {\frac 2{{4-{\rm D}}}}-\gamma _E+\ln (4\,\pi )\right) .}
\label{Fdivergent}
\end{eqnarray}
In these formulae ${\rm D}=4-2\epsilon $ and
\begin{equation}
J=\left\{ \matrix{ \sqrt{\tau_g-1},& \; \tau_g>1 \cr -i\sqrt{1-\tau_g},& \;
\tau_g<1}\right. .  \nonumber
\end{equation}

\section{Discussion}

Let us now briefly discuss how to interpret this result. As we have shown
above, the lagrangian (\ref{lagrangian}) is a mixture of terms stemming from
operators with different dimensions (from 6 up to 10 in the framework of the
decoupling scenario) and different orders in the momentum expansion (from $%
{\cal O}(p^2)$ up to ${\cal O}(p^6)$ in the framework of the nondecoupling
scenario). As a consequence, the formulas (\ref{Ffinite}), (\ref{Fdivergent}%
) do not respect the hierarchy of contributions originating from the
hierarchy of the tower of effective operators, which is the cornerstone of
the consistent treatment of the nonrenormalizable couplings within the
effective lagrangian approach. Therefore, in order to extract the partial
information about the relevant dependence on the scale of the new physics $%
\Lambda $, it is necessary to reorganize the resulting formulas and to keep
only the terms, which are dominant within the two possible scenarios
reviewed in the Sec. 1.

Within the framework of the decoupling scenario, we expect that we can
safely neglect\footnote{%
However,as it was shown in \cite{Artz}, in the case when the fundamental
full theory is weakly coupled gauge theory, the operators of dimension 8
should also contribute significantly provided they can be generated at the
tree level. In this case their contribution is comparable with that of the
one loop generated dimension 6 operators, which are multiplied by additional
factor $1/16\pi ^2$. This factor can be of the same size like the
suppression factor $v^2/\Lambda ^2$ of the dimension 8 operators, provided
the scale of new physics is in the range of a few TeV.\label{dimsix}} the
operators of dimension 8 and higher and keep only operators of dimension 6.
The lagrangian (\ref{decoupling}) contains two such operators (we use here
the notation of \cite{HISZ}), namely

\begin{eqnarray}
{\cal O}_B &=&\frac{{\rm i}g^{^{\prime }}}2B_{\mu \nu }D^\mu \Phi ^{+}D^\nu
\Phi ,  \nonumber \\
{\cal O}_{WWW} &=&-\frac{g^3}{3!}{\rm tr}(\widehat{W}_{\nu \mu }\widehat{W}%
^{\mu \rho }\widehat{W}_\rho ^{\;\;\nu }),  \label{OBOWWW}
\end{eqnarray}
so that we can use (\ref{Ffinite},\ref{Fdivergent}) to get information about
the contribution of these two operators only.

As far as the $WW\gamma $ and $WW\gamma \gamma $ couplings used for the
above calculation of the decay amplitude are concerned, the presence of the
term $(\alpha _B/\Lambda ^2)\,{\cal O}_B$ in the lagrangian generates
effectively a contribution to the phenomenological parameter $\Delta \kappa
_\gamma $ (and not to the $\lambda _\gamma $)

\begin{equation}
\Delta \kappa _\gamma ^B=\frac 12\frac{m_W^2}{\Lambda ^2}\alpha _B,
\label{kappaB}
\end{equation}
while the term $(\alpha _{WWW}/\Lambda ^2)\,{\cal O}_{WWW}$ contributes to
the $\lambda _\gamma $ (and not to the $\Delta \kappa _\gamma $)

\begin{equation}
\lambda _\gamma ^{WWW}=-\frac 14g^2\frac{m_W^2}{\Lambda ^2}\alpha _{WWW}.
\label{lambdaWWW}
\end{equation}
I.e., within the decoupling scenario, the natural values of the parameters $%
\Delta \kappa _\gamma $ and $\lambda _\gamma $ are of the order ${\cal O}(%
\frac{m_W^2}{\Lambda ^2})$ and ${\cal O}(g^2\frac{m_W^2}{\Lambda ^2})$
respectively\footnote{%
Let us stress that within the full $SU(2)\times U(1)$ approach, the operator
${\cal O}_B$ would induce also the anomalous $HWW\gamma $ coupling; however
within the lagrangian (\ref{decoupling}) this contribution is cancelled by
the contributions of the higher dimension operators with unnaturally large
coupling constants. That is, the lagrangian (\ref{decoupling}) does not
allow to get the full information on the influence of the operator ${\cal O}%
_B$ on the process $H\rightarrow \gamma \gamma $ using the above formulas (%
\ref{Ffinite},\ref{Fdivergent}), which is restricted to the effect of the
anomalous $WW\gamma $ and $WW\gamma \gamma $ vertices generated by ${\cal O}%
_B$ . On the other hand, the contribution of the loops with one insertion of
the anomalous $WW\gamma $ and $WW\gamma \gamma $ vertices generated by
dimension 6 operators (\ref{OBOWWW}) can be inferred therefore from (\ref
{Ffinite},\ref{Fdivergent}) by means of the expansion to the order ${\cal O}%
(\Delta \kappa _\gamma ,\lambda _\gamma ).$} and the leading order anomalous
contribution is therefore

\begin{eqnarray}
F_g^{fin} &=&2+3\tau _g+3\tau _g\left( 2-\tau _g\right) I^2  \nonumber \\
&&\ \ \ \ -2\Delta \kappa _\gamma +8\left( \lambda _\gamma +\Delta \kappa
_\gamma \frac 1{\tau _g}\right) -4\lambda _\gamma \tau _gI^2  \nonumber \\
&&\ \ \ \ +4\left[ 4\lambda _\gamma -3\Delta \kappa _\gamma +2\Delta \kappa
_\gamma \frac 1{\tau _g}+(\Delta \kappa _\gamma ^B-4\lambda _\gamma )\tau
_g\right] \frac IJ  \nonumber \\
&&\ \ \ \ -4(3\lambda _\gamma +\Delta \kappa _\gamma ^B\frac 1{\tau _g})\ln
\left( \frac{m_W{^2}}{{{{\mu }^2}}}\right) +{\cal O}\left( \frac 1{\Lambda
^4}\right)   \label{FOBOWWWfin} \\
F_g^{div} &=&-4\left( 3\lambda _\gamma +\Delta \kappa _\gamma \frac 1{\tau
_g}\right) {\left( {\frac 2{{4-{\rm D}}}}-\gamma _E+\ln (4\,\pi )\right) .}
\label{FOBOWWWdiv}
\end{eqnarray}

According to the renormalization prescription for the decoupling scenario,
the divergent part should be cancelled by the contributions of the
appropriate counterterms stemming from effective operators of dimension 6.
In the unitary gauge, such a counterterm has the form\footnote{%
Here we explicitly factored out the suppression factor $v^2/\Lambda ^2$ of
the dimension 6 operators. There could be also additional suppression of the
order $1/(4\pi )^2$ for the one loop generated dimension 6 operators.}

\begin{equation}
{\cal L}_{H\gamma \gamma }=G_{H\gamma \gamma }\frac{v^2}{\Lambda ^2}%
e^2\left( \frac Hv\right) F_{\mu \nu }F^{\mu \nu },  \label{counterterm}
\end{equation}
where $G_{H\gamma \gamma }$ is an effective (running) coupling constant%
\footnote{%
This form of interaction is generated e.g. by fixing the unitary gauge in
the following dimension 6 operators
\par
\begin{eqnarray}
{\cal O}_{WW} &=&g^2\Phi ^{+}\widehat{W}_{\mu \nu }\widehat{W}^{\mu \nu
}\Phi ,  \nonumber \\
{\cal O}_{BB} &=&-\frac{g^{^{\prime }2}}4\Phi ^{+}B_{\mu \nu }B^{\mu \nu
}\Phi ,  \nonumber \\
{\cal O}_{BW} &=&-\frac{{\rm i}gg^{^{\prime }}}2\Phi ^{+}B_{\mu \nu }%
\widehat{W}^{\mu \nu }\Phi .
\end{eqnarray}
Writing the corresponding terms of the effective lagrangian in the form
\par
\begin{equation}
{\cal L}_{ct}=\frac{\alpha _{WW}}{\Lambda ^2}{\cal O}_{WW}+\frac{\alpha _{BB}%
}{\Lambda ^2}{\cal O}_{BB}+\frac{\alpha _{BW}}{\Lambda ^2}{\cal O}_{BW}
\label{ctoperators}
\end{equation}
we have\footnote{%
Tree level generated dimension 8 operators could also give significant
contribution to $G_{H\gamma \gamma }$, cf. footnote \ref{dimsix} and \cite
{mex2}.}
\par
\begin{equation}
G_{H\gamma \gamma }=-\frac 14(\alpha _{WW}+\alpha _{BB}-\alpha _{BW}),
\end{equation}
Let us also note, that the operators (\ref{ctoperators}) lead to the gauge
boson wave function renormalization and mixing, as well as to the anomalous $%
HWW\gamma $ and $HWW\gamma \gamma $ vertices, which should also be included
in the complete analysis of the process under consideration, however these
effects are not discussed here.}. This brings about the following additional
contribution to the function $F_g$ :
\begin{equation}
F_g^{H\gamma \gamma }=\frac{(4\pi v)^2}{\Lambda ^2}G_{H\gamma \gamma }.
\label{treedecoupling}
\end{equation}
.

The above result of the loop calculation can be also used to get information
on the scale dependence of the effective constant $G_{H\gamma \gamma }.$ In
order to ensure the scale independent result for the decay rate it should
hold\footnote{%
Within the full $SU(2)\times U(1)$ approach, the ellipses here would stand
for the other terms coming from the other graphs with vertices generated by $%
{\cal O}_B$ as well as from other dimension 6 operators not considered here.}

\begin{eqnarray}
G_{H\gamma \gamma }(\mu ^{\prime }) &=&G_{H\gamma \gamma }(\mu )-\frac{%
\Lambda ^2}{(4\pi v)^2}8\left( 3\lambda _\gamma +\Delta \kappa _\gamma \frac
1{\tau _g}\right) \ln \left( \frac{\mu ^{\prime }}{{\mu }}\right) +\ldots
\nonumber
\end{eqnarray}

Within the framework of the nondecoupling scenario, the lowest order
anomalous contribution can be obtained by keeping only the contribution of
the ${\cal O}(p^4)$ operators. There are three such operators in the
lagrangian (\ref{nondecoupling}). The remaining three operators are of order
${\cal O}(p^6)$, these operators are proportional to the parameters $\lambda
_V.$ I.e. setting $\lambda _\gamma \rightarrow 0$ in the formulas (\ref
{Ffinite},\ref{Fdivergent}) and expanding to the first order in $\Delta
\kappa _\gamma $ we get the leading order contribution to the decay
amplitude from the very specific combination of the ${\cal O}(p^4)$
operators (here we use the same notation as in \cite{Longhitano})

\begin{eqnarray}
{\cal L}_2 &=&{\rm i}\frac{g^{^{\prime }}}2\ B_{\mu \nu }{\rm tr}(T[V^\mu
,V^\nu ])  \nonumber \\
{\cal L}_4 &=&\ [{\rm tr}(V_\mu V_\nu )]^2  \nonumber \\
{\cal L}_5 &=&\ [{\rm tr}(V_\mu V^\mu )]^2  \nonumber \\
{\cal L}_6 &=&\ {\rm tr}(V_\mu V_\nu ){\rm tr}(TV^\mu ){\rm tr}(TV^\nu )
\nonumber \\
{\cal L}_7 &=&\ {\rm tr}(V_\mu V^\mu ){\rm tr}(TV_\nu ){\rm tr}(TV^\nu ))
\nonumber \\
{\cal L}_9 &=&\ {\rm i}\frac g2{\rm tr}(T\widehat{W}_{\mu \nu }){\rm tr}%
(T[V^\mu ,V^\nu ]),  \label{L2-9}
\end{eqnarray}
with coefficients given by (\ref{nondecoupling}), cf. also footnote \ref
{footnote}. The trilinear vector boson coupling is generated by the
operators ${\cal L}_2$ and ${\cal L}_9$. Explicitly, the presence of the
terms
\begin{equation}
{\cal L}=\alpha _2\frac{v^2}{\Lambda ^2}{\cal L}_2+\alpha _9\frac{v^2}{%
\Lambda ^2}{\cal L}_9
\end{equation}
in the effective lagrangian give rise to the following contribution to the
parameter $\Delta \kappa _\gamma $:
\begin{equation}
\Delta \kappa _\gamma =-g^2\frac{v^2}{\Lambda ^2}(\alpha _2+\alpha _9)=-4%
\frac{m_W^2}{\Lambda ^2}(\alpha _2+\alpha _9),  \label{kappa29}
\end{equation}
i.e. the natural value for the coupling $\Delta \kappa _\gamma $ of the
lagrangian (\ref{nondecoupling}) is of the order ${\cal O}(\frac{m_W^2}{%
\Lambda ^2})$. For the leading order contribution to the function $F_g$ we
get then

\begin{eqnarray}
F_g^{fin} &=&2+3\tau _g+3\tau _g\left( 2-\tau _g\right) I^2  \nonumber \\
&&\ \ \ \ +2\Delta \kappa _\gamma {\biggl( }-1+4\frac 1{\tau _g}+2\left[
-3+2\frac 1{\tau _g}+\tau _g\right] \frac IJ-2\frac 1{\tau _g}\ln \left(
\frac{m_W{^2}}{{{{\mu }^2}}}\right) {\biggr) }\ \   \nonumber \\
&&\ \ \ \ +{\cal O}\left( \frac 1{\Lambda ^4}\right) \ \
\end{eqnarray}

\begin{equation}
F_g^{div}=-4\Delta \kappa _\gamma \frac 1{\tau _g}{\left( {\frac 2{{4-{\rm D}%
}}}-\gamma _E+\ln (4\,\pi )\right) .}  \label{chiraldiv}
\end{equation}
Note, that the divergent part is proportional to the $m_H^2=(k+l)^2$, this
reflects the fact that in the nondecoupling case the divergencies should be
canceled by ${\cal O}(p^6)$ counterterms. In the unitary gauge such a
counterterm has the form\footnote{%
Here again the suppression factor $v^2m_H^2/\Lambda ^4$ corresponding to the
naive dimensional analysis was factored out. The natural value of the
symmetry breaking scale is $\Lambda \sim 4\pi v$.}

\begin{eqnarray}
{\cal L}_{H\gamma \gamma } &=&\widetilde{G}_{H\gamma \gamma }e^2\frac{%
v^2m_H^2}{\Lambda ^4}\left( \frac Hv\right) F_{\mu \nu }F^{\mu \nu },
\end{eqnarray}
where $\widetilde{G}_{H\gamma \gamma }$ is an effective coupling constant
and we have

\begin{equation}
F_g^{H\gamma \gamma }=\frac{m_H^2}{\Lambda ^2}\widetilde{G}_{H\gamma \gamma
}.  \label{treenondecoupling}
\end{equation}
Also in this case we can infer information on the running of this effective
coupling; the scale dependence coming from the graphs with the above
mentioned specific combination of ${\cal O}(p^4)$ operators should be%
\footnote{%
Here, within the full nondecoupling approach, the ellipses would mean the
contributions of operators not listed above.}

\begin{eqnarray}
\widetilde{G}_{H\gamma \gamma }(\mu ^{\prime }) &=&\widetilde{G}_{H\gamma
\gamma }(\mu )-2\Delta \kappa _\gamma \left( \frac{\Lambda ^2}{4\pi vm_W}%
\right) ^2\ln \left( \frac{\mu ^{\prime }}{{\mu }}\right) +\ldots  \nonumber
\end{eqnarray}

\section{Conclusions}

Let us now compare these results with those of ref. \cite{Konig}. Inserting
the dimensional analysis estimates (\ref{kappaB}, \ref{lambdaWWW}, \ref
{kappa29}) of the parameters $\Delta \kappa _\gamma $ and $\lambda _\gamma $
within both scenarios to the cutoff analysis formula and using the principle
quoted in the Sec.2 (\ref{FKonig}), we get the following lower bound on the
dependence of the function $\Delta F_g$ on the scale of the new physics:

\begin{equation}
\Delta F_g^{{\rm cutoff}}=\frac 32\alpha _B+g^2\alpha _{WWW}+{\cal O}\left(
\frac{m_W^2}{\Lambda ^2}\right)  \label{Kdec}
\end{equation}
for the decoupling scenario and
\begin{equation}
\Delta F_g^{{\rm cutoff}}=-12(\alpha _2+\alpha _9)+{\cal O}\left( \frac{m_W^2%
}{\Lambda ^2}\right)  \label{Knon}
\end{equation}
for the nondecoupling scenario. I.e. in this scheme the natural values of
the anomalous contribution are (independently on the scale of the new
physics) ${\cal O}\left( 1\right) $ , provided the natural values of the LEC
$\alpha _B$, $\alpha _{WWW}$, $\alpha _2$ and $\alpha _9$ are of the order $%
{\cal O}\left( 1\right) $. On the other hand, within the dimensional
regularization approach the dominant anomalous contribution to the function $%
F_g$ comes from the direct interaction terms (cf. formulas (\ref
{treedecoupling}, \ref{treenondecoupling})) and the $\mu $ dependent loop
logarithms. For the purpose of the dimensional analysis, we can (using the
equations for the running of the LEC) interpret the scale $\mu $ as a scale
at which the constants $G_{H\gamma \gamma }$and $\widetilde{G}_{H\gamma
\gamma }$ acquire their natural values of order ${\cal O}\left( 1\right) $.
This is expected to correspond to the point at which the underlying high
energy theory is matched with the low energy effective lagrangian, i.e. to
the scale of the new physics $\Lambda $. We have then the following estimate
\begin{eqnarray*}
\Delta F_g^{{\rm DR}} &=&\frac{(4\pi v)^2}{\Lambda ^2}G_{H\gamma \gamma
}-\left( \frac 12\frac{m_H^2}{\Lambda ^2}\alpha _B-3g^2\frac{m_W^2}{\Lambda
^2}\alpha _{WWW}\right) \ln \left( \frac{m_W{^2}}{{{{\Lambda }^2}}}\right)
+\ldots \\
&=&{\cal O}\left( \frac{(4\pi v)^2}{\Lambda ^2}\right) +{\cal O}\left( \frac{%
m_H^2}{\Lambda ^2}\ln \left( \frac{m_W{^2}}{{{{\Lambda }^2}}}\right) \right)
+\ldots
\end{eqnarray*}
for the decoupling case (there could be an overall factor $1/(4\pi )^2$ for
the loop generated dimension 6 operator contribution) and
\begin{eqnarray*}
\Delta F_g^{{\rm DR}} &=&\frac{m_H^2}{\Lambda ^2}\widetilde{G}_{H\gamma
\gamma }+4(\alpha _2+\alpha _9)\frac{m_H^2}{\Lambda ^2}\ln \left( \frac{m_W{%
^2}}{{{{\Lambda }^2}}}\right) +\ldots \\
&=&{\cal O}\left( \frac{m_H^2}{\Lambda ^2}\right) +{\cal O}\left( \frac{m_H^2%
}{\Lambda ^2}\ln \left( \frac{m_W{^2}}{{{{\Lambda }^2}}}\right) \right)
+\ldots
\end{eqnarray*}
for the nondecoupling case. The ellipses here mean further terms,
unimportant from the numerical point of view.

We can make the following conclusion.The above formulas show, that the
cutoff scheme is in disagreement with dimensional analysis approach in this
special case of the calculation of the process $H\rightarrow \gamma \gamma $%
. In the often considered cases (e.g. $\Lambda =1TeV$ for the decoupling
scenario and loop generated dimension 6 operators, or $\Lambda \leq 4\pi v$
and $m_H<\Lambda $ for the nondecoupling scenario) the formulas (\ref{Kdec}%
), (\ref{Knon}) overestimate the enhancement or suppression of the decay
rate of the process under consideration. Therefore, this explicit example
does not rank among the cases, which could be treated correctly within the
momentum cutoff prescription. This corresponds to the general expectations
expressed in the ref. \cite{Burgess}, where further examples of both the
failure and success of cutoff analysis within the unitary gauge can be found.

\bigskip
\noindent{\bf Acknowledgements}

We would like to thank to J. Ho\v{r}ej\v{s}\'{\i} for the encouraging
discussions and for careful reading of the manuscript and useful remarks on
it. This work has been supported in part by research grants GA\v{C}R-1460/95
and GAUK-166/95.

\begin{figure}[tbp]
\epsfig{file=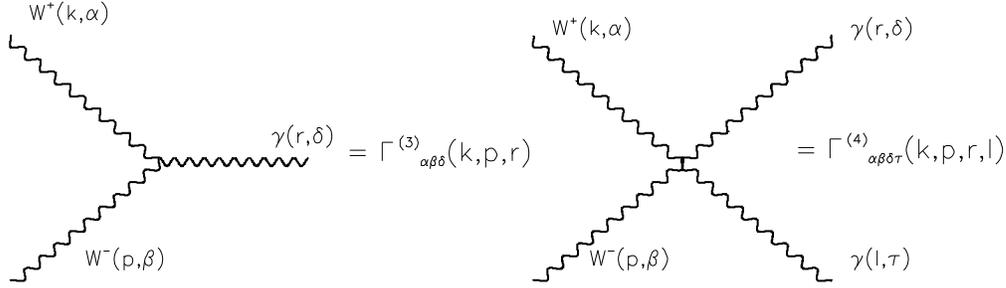}
\caption{Feynman rules for the $WW\gamma $ and $WW\gamma \gamma $ couplings.
All the momenta are in-going. The explicit expressions for the vertex
functions are given in the Appendix.}
\end{figure}

\section*{Appendix}

In this Appendix, we display some explicit formulas illustrating the
calculation of the decay rate $H\rightarrow \gamma \gamma $ within the
framework of the dimensional regularization. The $W$ -boson propagator in
the unitary gauge is given by the formula
\[
\Delta _F^{\mu \nu }(k)=-\frac{{\rm i}\left( g^{\mu \nu }-\frac{k_\mu k_\nu
}{m_W^2}\right) }{k^2-m_W^2+{\rm i}\varepsilon }
\]
and the Feynman rules for the $HWW$, $WW\gamma $ and $WW\gamma \gamma $
couplings are depicted in Fig. 1 The corresponding vertex functions are
\begin{eqnarray}
\Gamma _{\alpha \beta \delta }^{(3)}(k,p,r) &=&{\rm i}e\{g_{\alpha \beta
}(p-k)_\delta +g_{\alpha \delta }(k-r)_\beta +g_{\beta \delta }(r-p)_\alpha
\nonumber \\
&&+\Delta \kappa _\gamma (g_{\beta \delta }r_\alpha -g_{\alpha \delta
}r_\beta )  \nonumber \\
&&+\frac{\lambda _\gamma }{m_W^2}[(k\cdot p)(g_{\beta \delta }r_\alpha
-g_{\alpha \delta }r_\beta )+(r\cdot k)(g_{\alpha \beta }p_\delta -g_{\beta
\delta }p_\alpha )  \nonumber \\
&&+(r\cdot p)(g_{\alpha \delta }k_\beta -g_{\alpha \beta }k_\delta )+r_\beta
k_\delta p_\alpha -r_\alpha k_\beta p_\delta ]\}  \label{WWG}
\end{eqnarray}
and
\begin{eqnarray}
\Gamma _{\alpha \beta \delta \tau }^{(4)}(k,p,r,l) &=&-{\rm i}e^2(2g_{\alpha
\beta }g_{\delta \tau }-g_{\alpha \delta }g_{\beta \tau }-g_{\alpha \tau
}g_{b\delta }) \\
&&+\frac{{\rm i}e^2\lambda _\gamma }{m_W^2}\{-g_{\alpha \beta }g_{\delta
\tau }[(r+l)\cdot (k+p)]  \nonumber \\
&&+g_{\alpha \delta }g_{\beta \tau }[(r\cdot p)+(l\cdot k)]  \nonumber \\
&&+g_{\alpha \tau }g_{\beta \delta }[(r\cdot k)+(l\cdot p)]  \nonumber \\
&&+g_{\alpha \delta }[(k-p)_\tau r_\beta -r_\tau k_\beta -l_\beta k_\tau ]
\nonumber \\
&&+g_{\alpha \tau }[(k-p)_\delta l_\beta -r_\beta p_\delta -l_\delta k_\beta
] \\
&&+g_{\beta \delta }[-(k-p)_\tau r_\alpha -r_\tau p_\alpha -l_\alpha p_\tau ]
\nonumber \\
&&+g_{\beta \tau }[-(k-p)_\delta l_\alpha -l_\delta p_\alpha -r_\alpha
p_\delta ]  \nonumber \\
&&+g_{\alpha \beta }[(k+p)_\delta r_\tau +(k+p)_\tau l_\delta ]  \nonumber \\
&&+g_{\delta \tau }[k_\beta (r+l)_\alpha +p_\alpha (r+l)_\beta ]\}.
\label{WWGG}
\end{eqnarray}

\begin{figure}[tbp]
\epsfig{file=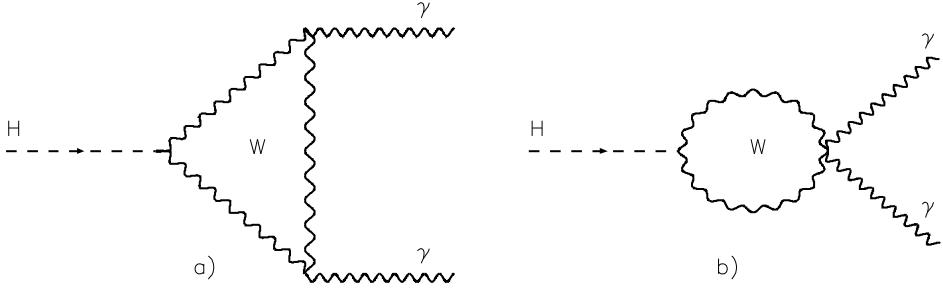}
\caption{The triangle and bubble graphs contributing to one loop amplitude
of the decay $H\rightarrow \gamma \gamma $.}
\end{figure}
In these expressions, the first rows represent the standard model vertices.
The decay amplitude is then given by the formula
\begin{equation}
{\cal M}(H\rightarrow \gamma \gamma )=\varepsilon ^{\star \mu
_1}(k_1)\varepsilon ^{\star \mu _2}(k_2){\cal M}_{\mu _1\mu _2},
\label{eq135}
\end{equation}
where $k_{1,2}$ are the momenta of the out-going photons and $\varepsilon $%
's are their polarization vectors. The polarization tensor ${\cal M}_{\mu
_1\mu _2}$ can be splitted into two parts
\begin{equation}
{\cal M}_{\mu _1\mu _2}={\cal M}_{\mu _1\mu _2}^a+{\cal M}_{\mu _1\mu _2}^b,
\label{eq136}
\end{equation}
where ${\cal M}_{\mu _1\mu _2}^a$ is the contribution of the triangle shown
in Fig. 2a (and correspondig crossed graph) and ${\cal M}_{\mu _1\mu _2}^b$
corresponds to the bubble in Fig. 2b, i.e.
\begin{eqnarray}
{\cal M}_{\mu _1\mu _2}^a &=&\int \frac{{\rm d}^{{\rm D}}l}{(4\pi )^{{\rm D}}%
}\frac{{\cal A}_{\mu _1\mu _2}^a(k_{1,}k_2,l)}{\left[ \left( l+k_1\right)
^2-m_W^2\right] \left[ \left( l-k_2\right) ^2-m_W^2\right] \left[
l^2-m_W^2\right] }  \nonumber  \label{eq137} \\
&&+\int \frac{{\rm d}^{{\rm D}}l}{(4\pi )^{{\rm D}}}\frac{{\cal A}_{\mu
_2\mu _1}^a(k_2,k_1,l)}{\left[ \left( l+k_2\right) ^2-m_W^2\right] \left[
\left( l-k_1\right) ^2-m_W^2\right] \left[ l^2-m_W^2\right] } \\
{\cal M}_{\mu _1\mu _2}^b &=&\int \frac{{\rm d}^{{\rm D}}l}{(4\pi )^{{\rm D}}%
}\frac{{\cal A}_{\mu _1\mu _2}^b(k_{1,}k_2,l)}{\left[ l^2-m_W^2\right]
\left[ \left( l-k_1-k_2\right) ^2-m_W^2\right] }.
\end{eqnarray}
Here

\begin{eqnarray}
{\cal A}_{\mu _1\mu _2}^a(k_{1,}k_2,l) &=&{\rm i}gm_W\left( g_{\alpha \mu }-%
\frac{(k_1+l)_\alpha (k_1+l)_\mu }{m_W^2}\right) \Gamma _{\mu \nu \mu
_1}^{(3)}(k_1+l,-l,-k_1)\left( g_{v\kappa }-\frac{l_\nu l_\kappa }{m_W^2}%
\right)  \nonumber \\
&&\times \Gamma _{\kappa \lambda \mu _2}^{(3)}(l,-l+k_2,-k_2)\left(
g_{\lambda \alpha }-\frac{(l-k_2)_\lambda (l-k_2)_\alpha }{m_W^2}\right)
\end{eqnarray}
and
\begin{eqnarray}
{\cal A}_{\mu _1\mu _2}^b(k_{1,}k_2,l) &=&{\rm i}gm_W\left( g_{\alpha \sigma
}-\frac{(k_1+k_2+l)_\alpha (k_1+k_2+l)_\sigma }{m_W^2}\right)  \nonumber \\
&&\times \Gamma _{\sigma \beta \mu _1\mu
_2}^{(4)}(k_1+k_2+l,-l,-k_1,-k_2)\left( g_{\beta \alpha }-\frac{l_\beta
l_\alpha }{m_W^2}\right) .
\end{eqnarray}
Using the standard Feynman parametrization and shifting the loop momenta we
get the following reprezentation, which allows for symmetric integration
over the loop momentum according to the standard formulas for the $D$%
-dimensional integration:
\begin{eqnarray}  \label{triangle}
{\cal M}_{\mu _1\mu _2}^a=2\int_0^1{\rm d}uu{\rm d}v\int \frac{{\rm d}^{{\rm %
D}}l}{(4\pi )^{{\rm D}}}\frac{{\cal A}_{\mu _1\mu
_2}^a(k_{1,}k_2,l-(uvk_1-u(1-v)k_2))+((k_{1,}\mu _1)\leftrightarrow
(k_{2,}\mu _2))}{(l^2-C^a(u,v))^3},  \nonumber \\
\end{eqnarray}
where
\begin{equation}
C^a(u,v)=m_W^2-m_H^2u^2v(1-v)
\end{equation}
and
\begin{equation}
{\cal M}_{\mu _1\mu _2}^b=\int_0^1{\rm d}u\int \frac{{\rm d}^{{\rm D}}l}{%
(4\pi )^{{\rm D}}}\frac{{\cal A}_{\mu _1\mu _2}^b(k_{1,}k_2,l-u(k_1+k_2))}{%
(l^2-C^b(u))^2},  \label{bubble}
\end{equation}
where
\begin{equation}
C^b(u)=m_W^2+m_H^2u(1-u).
\end{equation}

The rest of the calculation ( {\it i.e.} expansion of the numerators of the
integrands (\ref{triangle}) and (\ref{bubble}), the symmetric integration
over the loop momenta, the integration over the Feynman parameters and
extraction of the finite and divergent parts) was performed using {\it \
Mathematica}. We also used the {\it Mathematica }package {\em FeynCalc} \cite
{feyncalc}, which proves to be extremely useful for this purpose.

\end{document}